\documentclass[english,aps,article,showpacs,showkeys]{revtex4}
\usepackage[T1]{fontenc}
\usepackage[latin9]{inputenc}
\usepackage{amssymb}
\usepackage{graphicx}
\usepackage{esint}

\makeatletter
\@ifundefined{textcolor}{}
{%
 \definecolor{BLACK}{gray}{0}
 \definecolor{WHITE}{gray}{1}
 \definecolor{RED}{rgb}{1,0,0}
 \definecolor{GREEN}{rgb}{0,1,0}
 \definecolor{BLUE}{rgb}{0,0,1}
 \definecolor{CYAN}{cmyk}{1,0,0,0}
 \definecolor{MAGENTA}{cmyk}{0,1,0,0}
 \definecolor{YELLOW}{cmyk}{0,0,1,0}
}

\makeatother

\usepackage{babel}
\begin{document}

\title{Confinement on $R^{3}\times S^{1}$: continuum and lattice}

\author{Michael C. Ogilvie}

\address{Washington University, St. Louis, MO 63130 USA}

\email{mco@physics.wustl.edu}

\date{03/06/14}
\begin{abstract}
There has been substantial progress in understanding confinement in
a class of four-dimensional SU(N) gauge theories using semiclassical
methods. These models have one or more compact directions, and much
of the analysis is based on the physics of finite-temperature gauge
theories. The topology $R^{3}\times S^{1}$ has been most often studied,
using a small compactification circumference $L$ such that the running
coupling $g^{2}\left(L\right)$ is small. The gauge action is modified
by a double-trace Polyakov loop deformation term, or by the addition
of periodic adjoint fermions. The additional terms act to preserve
$Z(N)$ symmetry and thus confinement. An area law for Wilson loops
is induced by a monopole condensate. In the continuum, the string
tension can be computed analytically from topological effects. Lattice
models display similar behavior, but the theoretical analysis of topological
effects is based on Abelian lattice duality rather than on semiclassical
arguments. In both cases the key step is reducing the low-energy symmetry
group from $SU(N)$ to the maximal Abelian subgroup $U(1)^{N-1}$
while maintaining $Z(N)$ symmetry.
\end{abstract}

\pacs{12.38.Aw, 11.15.Ha, 11.15.Kc, 11.10Wx}

\keywords{quark confinement, monopoles, duality}

\maketitle

\section{Introduction: gauge theories}

One of the fundamental aspects of QCD is quark confinement: the force
between widely separated quark-antiquark pairs is a constant $\sigma$,
known as the string tension. This constant force implies a potential
energy between a quark-antiquark pair that grows as $\sigma r$ for
large distances r. The numerical value of $\sigma$, determined from
phenomenology, is approximately $0.18\, GeV^{2}\approx0.9\, GeV/fm$.
Renormalization group arguments tell us that the string tension depends
on the coupling constant of QCD in a non-perturbative way: it cannot
be calculated from perturbation theory. It is often convenient theoretically
to view QCD in a simplifed way, in a form without dynamical
quarks; this is often referred to as pure $SU(3)$ gauge theory. This
is a theory with no adjustible dimensionless parameters. As a consequence
of dimensional transmutation, the dimensionless gauge coupling $g^{2}$
can be replaced by a parameter $\Lambda$ with dimensions of energy.
Observables become pure numbers times an appropriate power of $\Lambda$.
The pure gauge theory is thus a theory with no adjustible dimensionless
parameters, making it difficult to carry out analytical approximations.
Finite-temperature gauge theories offer a dimensionless parameter
$T/\Lambda$ which can be used, for example, to change QCD from confining
behavior at low temperatures, with a non-zero string tension, to deconfined
behavior at high temperatures with zero string tension. Thus temperature
allows us to probe the physics of confinement. Recent work has shown
the existence of a new class of gauge theory models which provide
an analytic understanding of confinement in a class of four-dimensional
gauge theories. All of these models have one or more compact directions,
and the most developed case is the geometry $R^{3}\times S^{1}$,
which is the geometry of Euclidean gauge theories at finite temperature
when the circumference $L$ of $S^{1}$ is identified with the inverse
temperature $\beta=1/T$. Unlike conventional finite-temperature gauge
theories, this new class can be put into a confined phase when $L\ll\Lambda$
and $g^{2}\left(L\right)\ll1$ \cite{Unsal:2007vu}. Euclidean monopoles,
the constituents of finite-temperature instantons, are essential to
a semiclassical calculation of the string tension in this region,
explicitly revealing the nonperturbative nature of the string tension.
Moreover, this small-$L$ phase is smoothly connected to the conventional,
large-$L$ confining phase \cite{Myers:2007vc}.

In the next three sections, the basic features of these confining
models are developed for continuum quantum field theory. The simplest
case of the gauge group $SU(2)$ is used throughout as an example.
More details, as well as an introduction to some related topics, can
be found in \cite{Ogilvie:2012is}. The next section introduces some
basic ideas about confinement and center symmetry in finite-temperature
gauge theories. Section III show how the gauge action can be modified
to maintain confinement at small $L$. Section IV examines the nonperturbative
content of these models in the continuum and shows how topological
effects produce a nonzero string tension. Section V explores the nonperturbative
content of these models on the lattice, showing a close relation between
the continuum and the lattice. A final section summarizes. The notations
used throughout are as follows: All field theories are taken to be
in Euclidean space unless noted otherwise. Lower-case Greek indices
are used for space-time and the metric in Euclidean space is$g_{\mu\nu}=g^{\mu\nu}=\delta_{\mu\nu}.$
Roman indices in the range $j\cdots n$ generally denote ``spatial''
directions on $R^{3}\times S^{1}$, i.e., the three directions orthogonal
to the compact direction. Roman indices in the range $a\cdots d$
generally label group generators, while capital letters are used to
denote group representations: $F,Adj,S,A,R$. $S^{k}$ is the $k$-dimensional
surface of a $\left(k+1\right)$-dimensional hypersphere, so $S^{1}$
is the unit circle. $T^{k}$ is the $k$-dimensional hypertorus, so
$T^{1}$ is also $S^{1}$.

\section{\label{sec:Symmetries-of-gauge-theories}Confinement and $Z(N)$
symmetry}

Gauge theories with $T\ne0$ (``finite temperature'') have a rich
phase structure which has been extensively explored using a combination
of analytic methods and lattice simulations. Non-Abelian gauge theories
have global symmetries and associated order parameters which are analogous
to magnetization in spin systems, and much of the modern formalism
of critical phenomena is directly applicable. The symmetries and order
parameters associated with quark confinement and chiral symmetry breaking
are of particular interest as principal determinants of gauge theory
phase structure.

\begin{figure}
\includegraphics[scale=0.4]{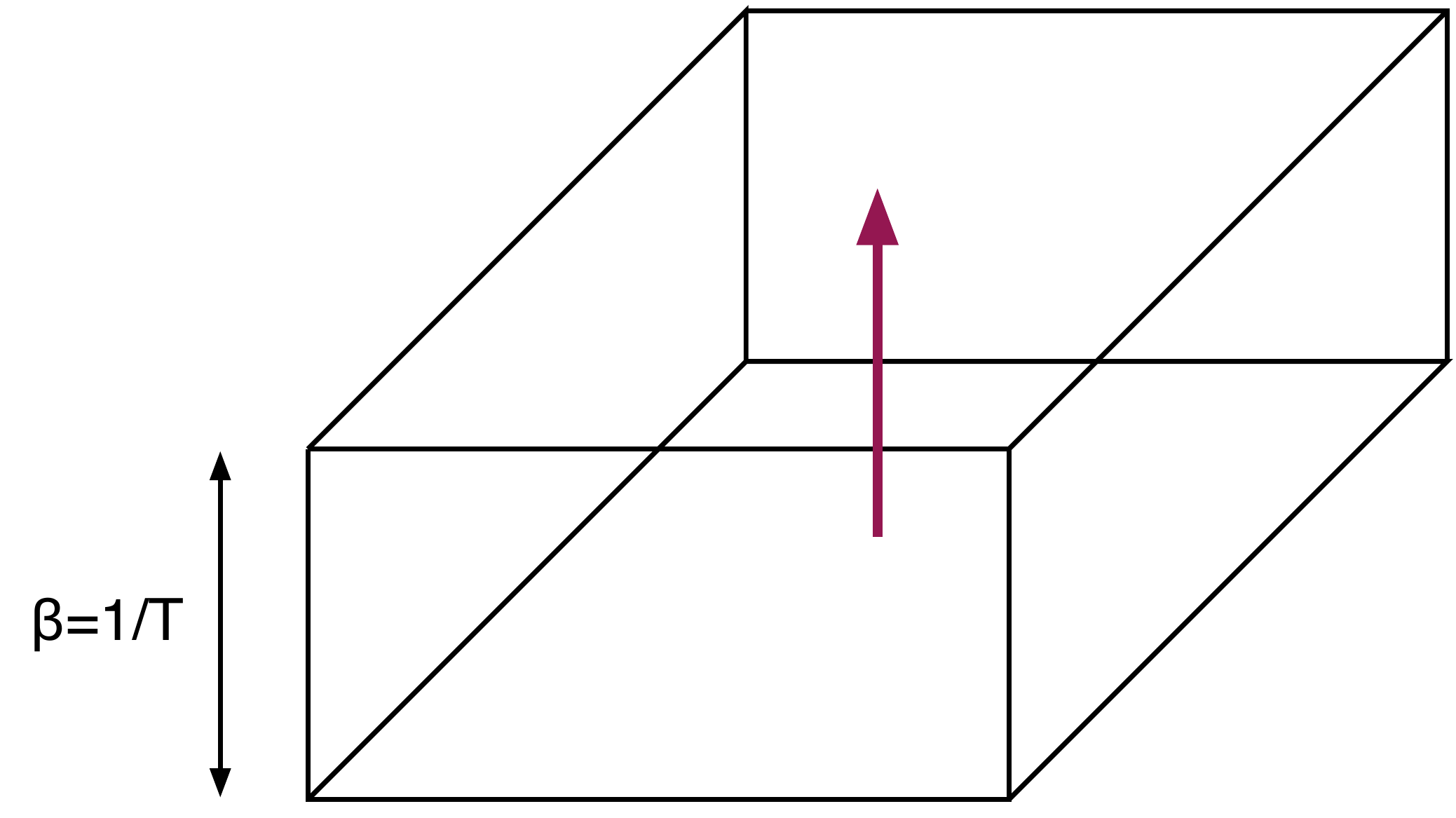}\caption{\label{fig:The-Polyakov-loop}The Polyakov loop is associated with
the worldline of a heavy particle.}
\end{figure}

\begin{figure}
\includegraphics[scale=0.4]{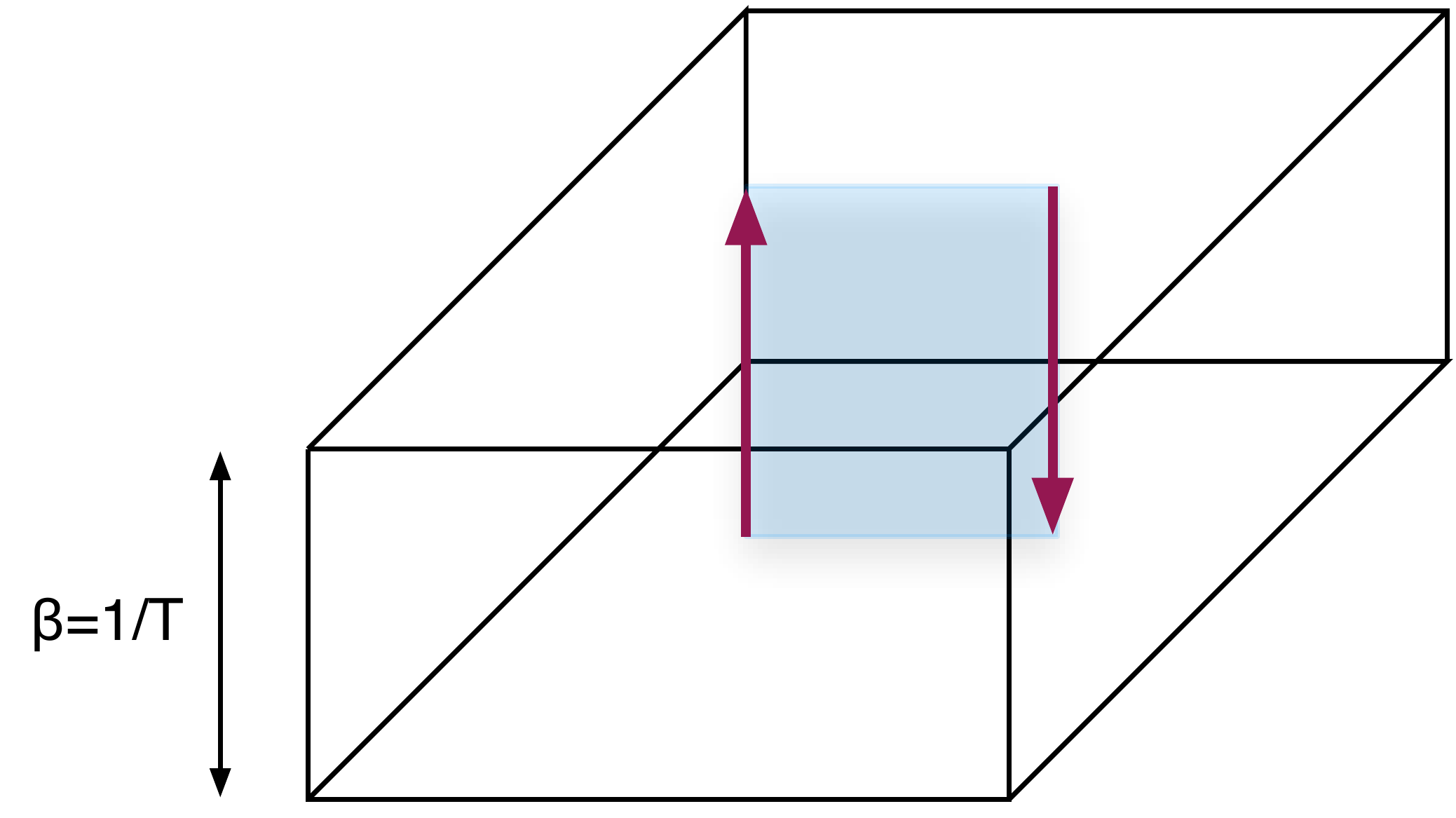}\caption{\label{fig:Polyakov-loop-2pt}The Polyakov loop two-point function
determines the electric string tension at finite temperature.}
\end{figure}

If one or more directions in space-time are compact, the string tension
may be measured using the Polyakov loop $P$, also known as the Wilson
line. The Polyakov loop is essentially a Wilson loop that uses a compact
direction in space-time to close the curve using a topologically non-trivial
path in space time, as shown in Figure \ref{fig:The-Polyakov-loop}.
The typical use for the Polyakov loop is for gauge theories at finite
temperature, where space-time is $R^{3}\times S^{1}$. The partition
function is given by $Z=Tr\left[e^{-\beta H}\right]$, with the circumference
of $S^{1}$ given by the inverse temperature $\beta=1/T$. In this
case, we write
\begin{equation}
P\left(\vec{x}\right)=\mathcal{P}\exp\left[i\int_{0}^{\beta}dx_{4}A_{4}\left(x\right)\right]
\end{equation}
where $\mathcal{P}$ indicates path-ordering of the integral. The
Polyakov loop one-point function $\left\langle Tr_{R}P\left(\vec{x}\right)\right\rangle $
can be interpreted as a Boltzman factor $\exp\left(-\beta F_{R}\right)$,
where $F_{R}$ is the free energy required to add a static particle
in the representation $R$ to the system. Of course, $\left\langle Tr_{R}P\left(\vec{x}\right)\right\rangle =0$
implies that $F_{R}=\infty$ , which is thus a fundamental criterion
determining whether particles in the representation $R$ are confined.
If $\left\langle Tr_{R}P\left(\vec{x}\right)\right\rangle =0$ , the
string tension associated with $R$ may be determined from a two-point
function 
\begin{equation}
\left\langle Tr_{R}P\left(\vec{x}\right)Tr_{R}P^{\dagger}\left(\vec{y}\right)\right\rangle \sim e^{-\beta\sigma_{R}\left|\vec{x}-\vec{y}\right|}
\end{equation}
as shown in Figure \ref{fig:Polyakov-loop-2pt}. Note that the introduction
of a compact direction breaks the four-dimensional symmetry of the
theory, and the string tension measured by Polyakov loops is not the
same as the string tension measured by Wilson loops lying in non-compact
planes. In the case of finite temperature, it is natural to use the
terminology electric and magnetic string tension, respectively. In
the limit where the compactification radius becomes large, \emph{i.e.}
$\beta\rightarrow\infty$, the two string tensions must coincide.

One of the most important concepts in our understanding of confinement
is the role of center symmetry. The center of a Lie group is the set
of all elements that commute with every other element. For $SU(N)$,
this is $Z(N)$. Although the $Z(N)$ symmetry of $SU(N)$ gauge theories
can be understood from the continuum theory, it is easier to understand
from a lattice point of view. A lattice gauge theory associates link
variable $U_{\mu}\left(x\right)$ with each lattice site $x$ and
direction $\mu$. The link variable is considered to be the path-ordered
exponential of the gauge field from $x$ to $x+\hat{\mu}$: $U_{\mu}\left(x\right)=\exp\left[iaA_{\mu}\left(x\right)\right].$
Consider a center symmetry transformation on all the links in a given
direction on a fixed hyperplane perpendicular to the direction. The
standard example from $SU(N)$ gauge theories at finite temperature
is $U_{4}\left(\vec{x},t\right)\rightarrow zU_{4}\left(\vec{x},t\right)$
for all $\vec{x}$ and fixed $t$, with $z\in Z(N)$. Because lattice
actions such as the Wilson action consist of sums of small Wilson
loops, they are invariant under this global symmetry. However, the
Polyakov loop transforms as $P\left(\vec{x}\right)\rightarrow zP\left(\vec{x}\right)$,
and more generally 
\begin{equation}
Tr_{R}P\left(\vec{x}\right)\rightarrow z^{k_{R}}Tr_{R}P\left(\vec{x}\right)
\end{equation}
 where $k_{R}$ is an integer in the set $\left\{ 0,1,...,N-1\right\} $
and is known as the $N$-ality of the representation $R$. If $k_{R}\ne0$,
then unbroken global $Z(N)$ symmetry implies $\left\langle Tr_{R}P\left(\vec{x}\right)\right\rangle =0$.
Thus global $Z(N)$ symmetry characterizes the confining phase of
an $SU(N)$ gauge theory. For pure gauge theories at non-zero temperature,
the deconfinment phase transition is associated with the loss of $Z(N)$
symmetry at the critical point $T_{d}$. Below that point $\left\langle Tr_{F}P\left(\vec{x}\right)\right\rangle =0$
but above $T_{d}$, $\left\langle Tr_{R}P\left(\vec{x}\right)\right\rangle \ne0$.
Notice that the case of zero $N$-ality representations is special
within this framework: there is no requirement from $Z(N)$ symmetry
that these representations are confined. This includes the adjoint
representation, the representation of the gauge particles. However,
lattice simulation indicate that $\left\langle Tr_{R}P\left(\vec{x}\right)\right\rangle $
is very small for these representations in the confined phase. Although
screening by gauge particles must dominate at large distances, these
zero $N$-ality representations have well-defined string tensions
at intermediate distances scales, \emph{e.g.}, on the order of a few
fermi for $SU(3)$, behaving in a manner very similar to representations
with non-zero $N$-ality \cite{Deldar:1999vi,Bali:2000un}.

In the confining phase of a gauge theory, $Z(N)$ symmetry is unbroken,
and all representations with non-zero $N$-ality are confined. In
the deconfined phase, $Z(N)$ symmetry is completely lost, and particles
are no longer confined, independent of their representation. For $N\ge4$,
additional phases are possible where $Z(N)$ is broken down to a non-trivial
subgroup \cite{Meisinger:2001cq,Myers:2007vc,Myers:2009df,Meisinger:2009ne}.
In the case of $Z(4)$, there can be breaking of center symmetry down
to $Z(2)$. In this partially confined phase, states consisting of
two fundamental representation fermions are not confined, but single
fermions are. In the case of $SU(N)$, $Z(N)$ can break to $Z(k)$,
where $k$ is any divisor of $N$. States with $1$ to $k-1$ fermions
are confined, but states with $k$ fermions are not. It is often convenient
to include the confined and deconfined phases as $k=N$ and $k=1$,
respectively. Such partially confined phases been found in gauge theories
on $R^{3}\times S^{1}$, using both lattice simulations and perturbation
theory \cite{Myers:2007vc}. It should be noted that not all gauge
groups have non-trivial centers. The gauge group $G(2)$ provides
an interesting example of a gauge theory without a center \cite{Holland:2003jy}
that has received significant attention \cite{Pepe:2005sz}.

The effective potential for the Polyakov loop should describe both
the confined and deconfined phases if perturbative and nonperturbative
effects are included. Perturbation theory is a reliable indicator
of broken center symmetry and thus deconfinement at high temperature,
because the running coupling constant $g(T)$ is small if $T\gg\Lambda$.
The one-loop effective potential for a pure gauge theory in the background
of a static Polyakov loop $P$ can be easily evaluated in a gauge
where the background field $A_{4}$ is time-independent and diagonal
\cite{Gross:1980br,Weiss:1980rj}. It is easy to see that $V_{eff}^{1l}\left(P\right)$
is given by
\begin{equation}
V_{eff}^{1l}\left(P\right)=2T\, Tr_{A}\sum_{n\in Z}\int\frac{d^{3}k}{\left(2\pi\right)^{3}}\log\left[\left(2\pi nT-A_{4}\right)^{2}+\vec{k}^{2}\right]
\end{equation}
where the factor of $2$ represents the two helicity states of each
mode. Note that there is no classical contribution to $V_{eff}^{1l}$.
Discarding the zero-point energy term, we obtain the one-loop finite-temperature
effective potential for gauge bosons
\begin{equation}
V_{eff}^{1l}\left(P\right)=2T\, Tr_{A}\int\frac{d^{3}k}{\left(2\pi\right)^{3}}\log\left[1-P\,\exp\left(-\left|\vec{k}\right|/T\right)\right]
\end{equation}
which is the free energy density of the gauge bosons in the background
$P$. The logarithm in this expression for $V_{eff}^{1l}\left(P\right)$
can be expanded, leading to an interpretation of $V_{eff}^{1l}\left(P\right)$
as a sum of contibutions from gluon worldlines wrapping around the
compact direction an arbitrary number of times. Explicitly, we have
the expression
\begin{equation}
V_{eff}^{1l}\left(P\right)=-\frac{2}{\pi^{2}}\sum_{n=1}^{\infty}\frac{1}{n^{4}}Tr_{A}P^{n}.
\end{equation}
From this form, it is easy to see that $ $$V_{eff}^{1l}\left(P\right)$
is minimized when all the moments $Tr_{A}P^{n}$ are maximized. This
occurs when $P\in Z(N)$, which gives $Tr_{A}P^{n}=N^{2}-1$. This
indicates that the one-loop gluon effective potential favors the deconfined
phase. The pressure $p$ is the negative of the free energy density
at the minimum, so 
\begin{equation}
p\left(T\right)=2\left(N^{2}-1\right)\frac{\pi^{2}T^{4}}{90},
\end{equation}
 which is exactly $p$ for a blackbody with $2\left(N^{2}-1\right)$
degrees of freedom. 

In the gauge where $A_{4}$ is diagonal and time-independent, we can
parametrize $A_{4}$ in the fundmental representation of $SU(N)$
as a diagonal, traceless $N\times N$ matrix
\begin{equation}
\left(A_{4}\right)_{jk}=T\theta_{j}\delta_{jk}
\end{equation}
so that
\begin{equation}
P_{jk}=e^{i\theta_{j}}\delta_{jk}
\end{equation}
with $\sum_{j=1}^{N}\theta_{j}=0$. Using the decomposition $F\otimes\bar{F}=1\oplus Adj$
and the corresponding decomposition$Tr_{F}P\, Tr_ {F} P^{+}=1+Tr_{Adj }P,$
one realizes that the $N^{2}$ eigenvalues of the product representation
$F\otimes\bar{F}$ have the form $\exp\left[i\Delta\theta_{jk}\right]$
, where we define $\Delta\theta_{jk}\equiv\theta_{j}-\theta_{k}$, giving
\begin{equation}
V_{eff}^{1l}\left(P\right)=-\frac{2}{\pi^{2}}\sum_{j,,k=1}^{N}(1-\frac{1}{N}\delta_{jk})\sum_{n=1}^{\infty}\frac{1}{n^{4}}\exp\left[in\Delta\theta_{jk}\right].
\end{equation}
The infinite sum over $n$ may be carried out explicitly in terms
of the fourth Bernoulli polynomial. For our purposes, a convenient
explicit form is 
\begin{equation}
V_{eff}^{1l}\left(P\right)=-T^{4}\sum_{j,k=1}^{N}(1-\frac{1}{N}\delta_{jk})\left[\frac{\pi^{2}}{45}-\frac{1}{24\pi^{2}}\left|\Delta\theta_{jk}\right|_{2\pi}^{2}\left(\left|\Delta\theta_{jk}\right|_{2\pi}-2\pi\right)^{2}\right]
\end{equation}
where $\left|\Delta\theta_{jk}\right|_{2\pi}$ lies in the interval
between 0 and $2\pi$ \cite{Gross:1980br}.

Confinement at low temperatures is nonperturbative in nature. In many
systems, broken symmetry phases are found at low temperatures and
symmetry is restored at high temperatures. The phase structure of
gauge theories as a function of temperature is unusual because the
broken-symmetry phase is the high-temperature phase. A lattice construction
of the effective action for Polyakov loops, valid for strong-coupling,
is instructive \cite{Polonyi:1982wz,Ogilvie:1983ss,Green:1983sd,Gross:1984wb}.
The spatial link variables may be integrated out exactly if spatial
plaquette interactions are neglected. Each spatial link variable then
appears only in two adjacent temporal plaquettes, and may be integrated
out exactly using the same techniques that are used in the Migdal-Kadanoff
real-space renormalization group \cite{Ogilvie:1983ss,Billo:1996pu}.
The resulting effective action has the form
\begin{equation}
S_{eff}=-\sum_{\left\langle jk\right\rangle }K\left[Tr_{F}P_{j}Tr_{F}P_{k}^{\dagger}+Tr_{F}P_{k}Tr_{F}P_{j}^{\dagger}\right]
\end{equation}
where $K$ is a function of the lattice gauge coupling $g^{2}$ and
the extent of the lattice in the Euclidean time direction $n_{t}$,
which is related to the temperature by $n_{t}a=1/T$. In the strong-coupling
limit of the underlying gauge theory, the explicit form for $K$ is
$K\simeq\left(1/g^{2}N\right)^{n_{t}}$ to leading order. In the weak-coupling
limit, a Migdal-Kadanoff bond-moving argument gives $K\simeq2N/g^{2}n_{t}$.
This effective action represents a $Z(N)$-invariant nearest-neighbor
interaction of a spin system where the Polyakov loops are the spins.
It depends only on gauge-invariant quantities. Standard expansion
techniques show that the $Z(N)$ symmetry is unbroken for small $K$,
and broken for $K$ large. This model explains why the high-temperature
phase of gauge theries is the symmetry-breaking phase: the relation
between $K$ and the underlying gauge theory parameters is such that
$K$ is small at low temperatures, and large at high temperatures,
exactly the reverse of a classical spin system where the coupling
is proportional to $T^{-1}$. For small values of $n_{t}$, the deconfinement
transition can be easily extracted, but the phase transition is in
the strong-coupling region and far from the continuum limit. A systematic
treatment of strong-coupling corrections has recently been shown to
yield values for the critical lattice couplings $\beta_{c}\equiv2N/g^{2}$
for $SU(2)$ and $SU(3)$ that are within a few percent of simulation
results for $4\le N_{t}\le16$ \cite{Langelage:2010yr}. However,
strong-coupling expansions typically have a finite radius of convergence,
and are inadequate to describe the complete phase diagram.

\section{\label{sec:Phases-on-R3xS1}Preserving $Z(N)$ symmetry on $R^{3}\times S^{1}$}

In the last few years, it has proven possible to construct four-dimensional
gauge theories for which confinement may be reliably demonstrated
using semiclassical methods valid for weak coupling \cite{Myers:2007vc,Unsal:2007vu}.
These models combine $Z(N)$ symmetry, the effective potential for
$P$, instantons, and monopoles into a satisfying picture of confinement
for a special class of models. All of the models in this class have
one or more small compact directions. Models with an $R^{3}\times S^{1}$
topology have been most investigated, and discussion here will focus
on this class. The circumference $L$ of $S^{1}$ is taken to be small,
\emph{i.e.}, $L\ll\Lambda^{-1}$ , so that $g(L)\ll1$ and perturbation
theory and semiclassical arguments are reliable. Because the $R^{3}\times S^{1}$
topology is natural at finite temperature, it is often useful to identify
$L$ with $\beta=1/T$ although that is not always possible. There
are two distinct aspects to the behavior of these models. First, the
action is modified in such a way that that center symmetry is maintained
for small $L$. Second, nonperturbative effects associated with finite-temperature
instantons and Euclidean-space monopoles are used to establish that
the string tension is nonzero. Thus we obtain a realization of a long-held
scenario for quark confinement, based on ideas originally proposed
by Mandelstam \cite{Mandelstam:1974vf,Mandelstam:1974pi} and 't Hooft
\cite{'tHooft:1977hy,'tHooft:1979uj}.

\begin{figure}
\includegraphics[scale=0.6]{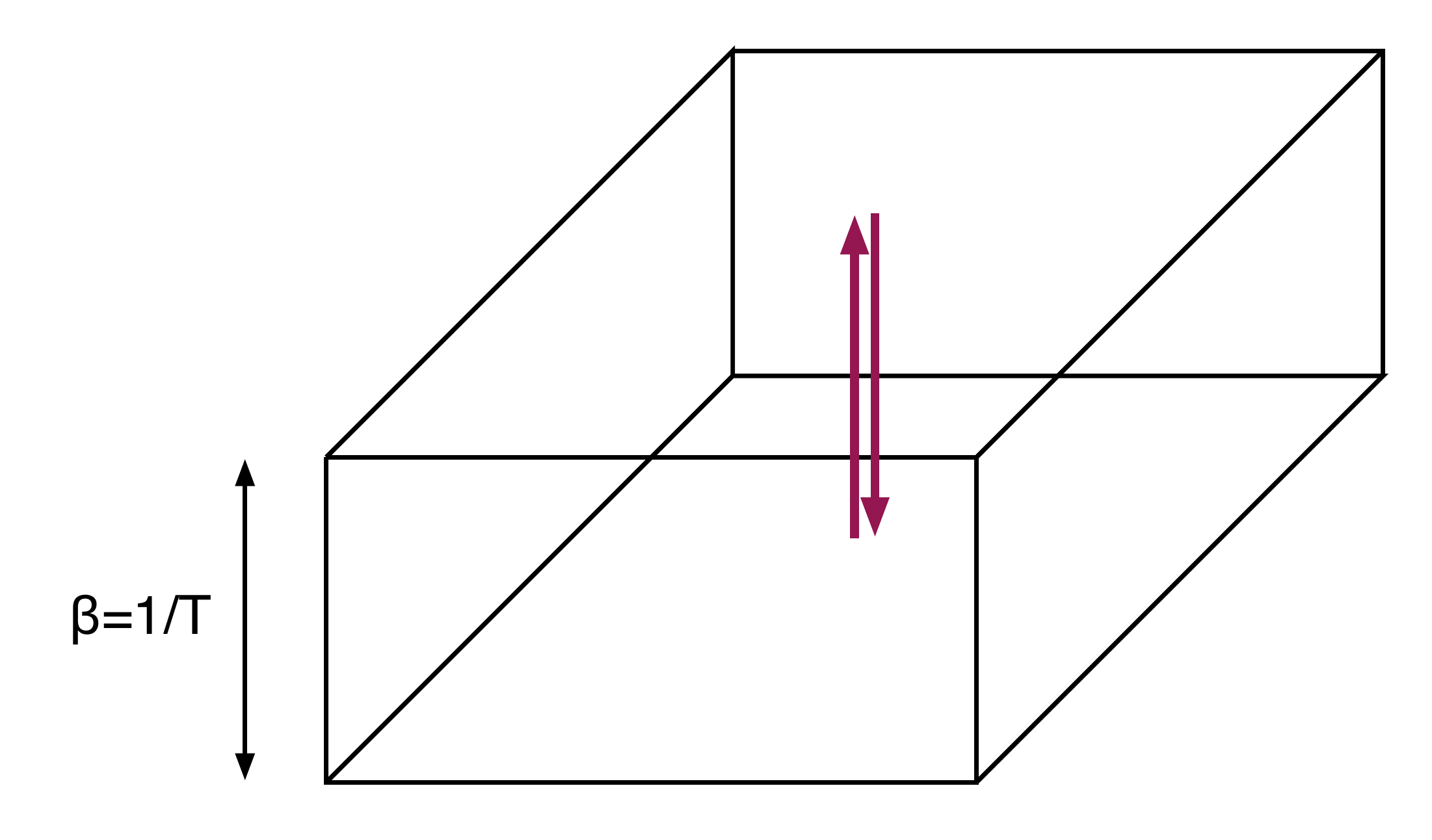}

\caption{\label{fig:double-trace}A double-trace Polyakov loop on $R^{3}\times S^{1}$.}
\end{figure}

There are two closely related approaches to maintaining $Z(N)$ symmetry
for small $L$. The first approach deforms the pure gauge theory by
adding additional terms involving the Polyakov loop to the gauge action
\cite{Myers:2007vc,Ogilvie:2007tj,Unsal:2008ch}. The general form
for such a deformation is
\begin{equation}
S\rightarrow S+\beta\int d^{3}x\,\sum_{k=1}^{\infty}a_{k}Tr_{A}P\left(\vec{x},x_{4}\right)^{k}
\end{equation}
where the value of $x_{4}$ is arbitrary and can be taken to be $0$.
Such terms are often referred to as double-trace deformations; see
Fig. \ref{fig:double-trace}. If the coefficients $a_{k}$ are sufficiently
large, they will counteract the effects of the one-loop effective
potential, and $Z(N)$ symmetry will hold for small $L$. Only the
first $\left[N/2\right]$ terms are necessary to ensure confinement.
It is easy to prove that for a classical Polyakov loop $P$, the conditions
$Tr_{F}P^{k}=0$ with $1\le k\le\left[N/2\right]$ determine the unique
set of Polyakov loop eigenvalues that constitute a confining solution,
\emph{i.e.}, one for which $Tr_{R}P=0$ for all representations with
$k_{R}\ne0$ \cite{Meisinger:2001cq}. The effective potential associated
with $S$ is given approximately by 
\begin{equation}
V_{eff}\left(P,\beta\right)=\frac{-2}{\pi^{2}\beta^{4}}\sum_{n=1}^{\infty}\frac{Tr_{A}P^{n}}{n^{4}}+\sum_{k=1}^{\left[\frac{N}{2}\right]}a_{k}Tr_{A}P^{k}.
\end{equation}
The explicit solution that minimizes the effective potential in the
confined phase is simple: up to a factor necessary to ensure $\det P=1$,
the eigenvalues of $P$ are given by the set of $N$'th roots of unity,
which are permuted by a global $Z(N)$ symmetry transformation. A
rich phase structure can emerge from the minimization of $V_{eff}$
for intermediate values of the coefficients $a_{k}$. For $N\ge3$,
the effective potential predicts that one or more phases may separate
the deconfined phase from the confined phase. In the case of $SU(3)$,
a single new phase is predicted, and has been observed in lattice
simulations \cite{Myers:2007vc}. For larger values of $N$, there
is a rich set of possible phases, including some where $Z(N)$ breaks
down to a proper subgroup $Z(p)$. In such phases, particles in the
fundamental representation are confined, but bound states of $p$
particles are not \cite{Ogilvie:2007tj}.

Lattice simulations of $SU(3)$ and $SU(4)$ agree for small $L$
with the theoretical predictions based on effective potential arguments
\cite{Myers:2007vc}. The phase diagram of $SU(3)$ as a function
of $T=L^{-1}$ and $a_{1}$ has three phases: the confined phase,
the deconfined phase, and a new phase, the skewed phase. In general,
the three phases of the eigenvalues of the Polykov loop may be taken
to be the set $\left\{ \theta_{1},\theta_{2},\theta_{3}\right\} $
where $\theta_{1}+\theta_{2}+\theta_{3}$=0. For all three phases,
it is possible to use $Z(3)$ symmetry to make $Tr_{F}P$ real, and
reduce the phases to the set $\left\{ 0,\theta,-\theta\right\} $
such that $Tr_{F}P=1+2\cos\theta$. The deconfined phase is represented
by $\theta=0$, the confined phase is given by $\theta=2\pi/3$, and
the skewed phase by $\theta=\pi.$ An important result obtained from
the lattice simulation of $SU(3)$ is that the small-$L$ confining
region, where semiclassical methods yield confinement, are smoothly
connected to the conventional large-$L$ confining region. In the
case of $SU(4)$, a sufficiently large value of $a_{1}$ leads to
a partially-confining phase where $Z(4)$ is spontaneously broken
to $Z(2)$. Particles with $k=1$ are confined in this phase, \emph{i.e.},
$\left\langle Tr_{F}P\left(\vec{x}\right)\right\rangle =0$, but particles
with $k=2$ are not, as indicated by $\left\langle Tr_{F}P^{2}\left(\vec{x}\right)\right\rangle \ne0$.
As in the case of $SU(3)$, $Tr_{F}P$ can be made real. Perturbation
theory then predicts a deconfined phase where the phases of the eigenvalues
of $P$ are $\left\{ 0,,0,0,0\right\} ,$ a confined phase where they
are $\left\{ \pi/4,3\pi/4,5\pi/4,7\pi/4\right\} $, and a partially
confined phase where the phases are $\left\{ \pi/2,\pi/2,-\pi/2,-\pi/2\right\} $.

Double-trace deformations may be applied to lattice gauge theories
as easily as to continuum models. A simple double-trace deformation
may be applied to the spin-model reduction of a finite-temperature
lattice gauge theory, resulting in an action of the form
\begin{equation}
S_{eff}=-K\sum_{\left\langle jk\right\rangle }\left[Tr_{F}P_{j}Tr_{F}P_{k}^{\dagger}+Tr_{F}P_{k}Tr_{F}P_{j}^{\dagger}\right]+A_{2}\sum_{j}Tr_{F}P_{j}Tr_{F}P_{j}^{\dagger}
\end{equation}
 where $A_{2}$ is the lattice analog of $a_{2}$ in the continuum.
In this form, it is clear that the deformation term acts directly
to oppose the tendency of the the nearest-neighbor interaction to
break $Z(N)$ symmetry.

Another approach to preserving $Z(N)$ symmetry for small $L$ uses
fermions in the adjoint representation with periodic boundary conditions
in the compact direction \cite{Unsal:2007vu}. In this case, it would
be somewhat misleading to use $\beta$ as a synonym for $L$, because
the transfer matrix for evolution in the compact direction is not
positive-definite. Periodic boundary conditions in the compact direction
imply that the generating function of the ensemble, \emph{i.e.}, the
partition function, is given by
\begin{equation}
Z=Tr\left[\left(-1\right)^{F}e^{-LH}\right]
\end{equation}
 where $F$ is the fermion number and $H$ is the Hamiltonian in the
compact direction. This graded ensemble, familiar from supersymmetry,
can be obtained from an ensemble $Tr\left[\exp\left(\beta\mu F-\beta H\right)\right]$
with chemical potential $\mu$ by the replacement $\beta\mu\rightarrow i\pi$.
This system can be viewed as a gauge theory with periodic boundary
conditions in one compact spatial direction of length $L=\beta$,
and the transfer matrix in the time direction is positive-definite.

The use of periodic boundary conditions for the adjoint fermions dramatically
changes their contribution to the Polyakov loop effective potential.
In perturbation theory, the replacement $\beta\mu\rightarrow i\pi$
shifts the Matsubara frequencies from $\beta\omega_{n}=\left(2n+1\right)\pi$
to $\beta\omega_{n}=2n\pi$. The one loop effective potential is now
essentially that of a bosonic field, but with an overall negative
sign due to fermi statistics \cite{Meisinger:2001fi}. The sum of
the effective potential for the fermions plus that of the gauge bosons
gives 
\begin{equation}
V_{eff}\left(P,\beta,m,N_{f}\right)=\frac{1}{\pi^{2}\beta^{4}}\sum_{n=1}^{\infty}\frac{Tr_{A}P^{n}}{n^{2}}\left[2N_{f}\beta^{2}m^{2}K_{2}\left(n\beta m\right)-\frac{2}{n^{2}}\right]\label{eq:Veff-adjoint-fermions-1}
\end{equation}
where $N_{f}$ is the number of adjoint Dirac fermions and $m$ is
their mass \cite{Meisinger:2009ne}. Note that the first term in brackets,
due to the fermions, is positive for every value of $n$, while the
second term, due to the gauge bosons, is negative. 

The largest contribution to the effective potential at high temperatures
is typically from the $n=1$ term, which can be written simply as
\begin{equation}
\frac{1}{\pi^{2}\beta^{4}}\left[2N_{f}\beta^{2}m^{2}K_{2}\left(\beta m\right)-2\right]\left[\left|Tr_{F}P\right|^{2}-1\right]
\end{equation}
 where the overall sign depends only on $N_{f}$ and $\beta m$. If
$N_{f}\ge1$ and $\beta m$ is sufficiently small, this term will
favor $Tr_{F}P=0$. On the other hand, if $\beta m$ is sufficiently
large, a value of $P$ from the center, $Z(N)$, is preferred. Note
that an $\mathcal{N}=1$ super Yang-Mills theory would correspond
to $N_{f}=1/2$ and $m=0$, giving a vanishing perturbative contribution
for all $n$ \cite{Davies:1999uw,Davies:2000nw}. In that case, non-perturbative
effects lead to a confining effective potential for all values of
$\beta$. In the case of $N_{f}\ge1$, each term in the effective
potential will change sign in succession as $m$ is lowered towards
zero. For larger values of $N$, this leads to a cascade of phases
separating the confined and deconfined phases \cite{Myers:2009df}.
Numerical investigation shows that the confined phase is obtained
if $N\beta m\lesssim4.00398$ \cite{Meisinger:2009ne}. As $m$ increases,
it becomes favorable that $Tr_{F}P^{n}\ne0$ for successive values
of $n$. If $N$ is even, the first phase after the confined phase
will be a phase with $Z(N/2)$ symmetry. As $m$ increases, the last
phase before reaching the deconfined phase will have $Z(2)$ symmetry,
in which $k=1$ states are confined, but all states with higher $k$
are not. Lattice simulations of $SU(3)$ with periodic adjoint fermions
are completely consistent with the picture \cite{Cossu:2009sq} predicted
by the effective potential, with a skewed phase separating the confined
phase and deconfined phase. For $N\ge3$, there are generally phases
intermediate between the confined and deconfined phases which are
not of the partially-confined type. Careful numerical analysis is
necessary on a case-by-case basis to determine the phase structure
for each value of $N$ \cite{Myers:2009df}. 

\begin{figure}
\includegraphics[width=5in]{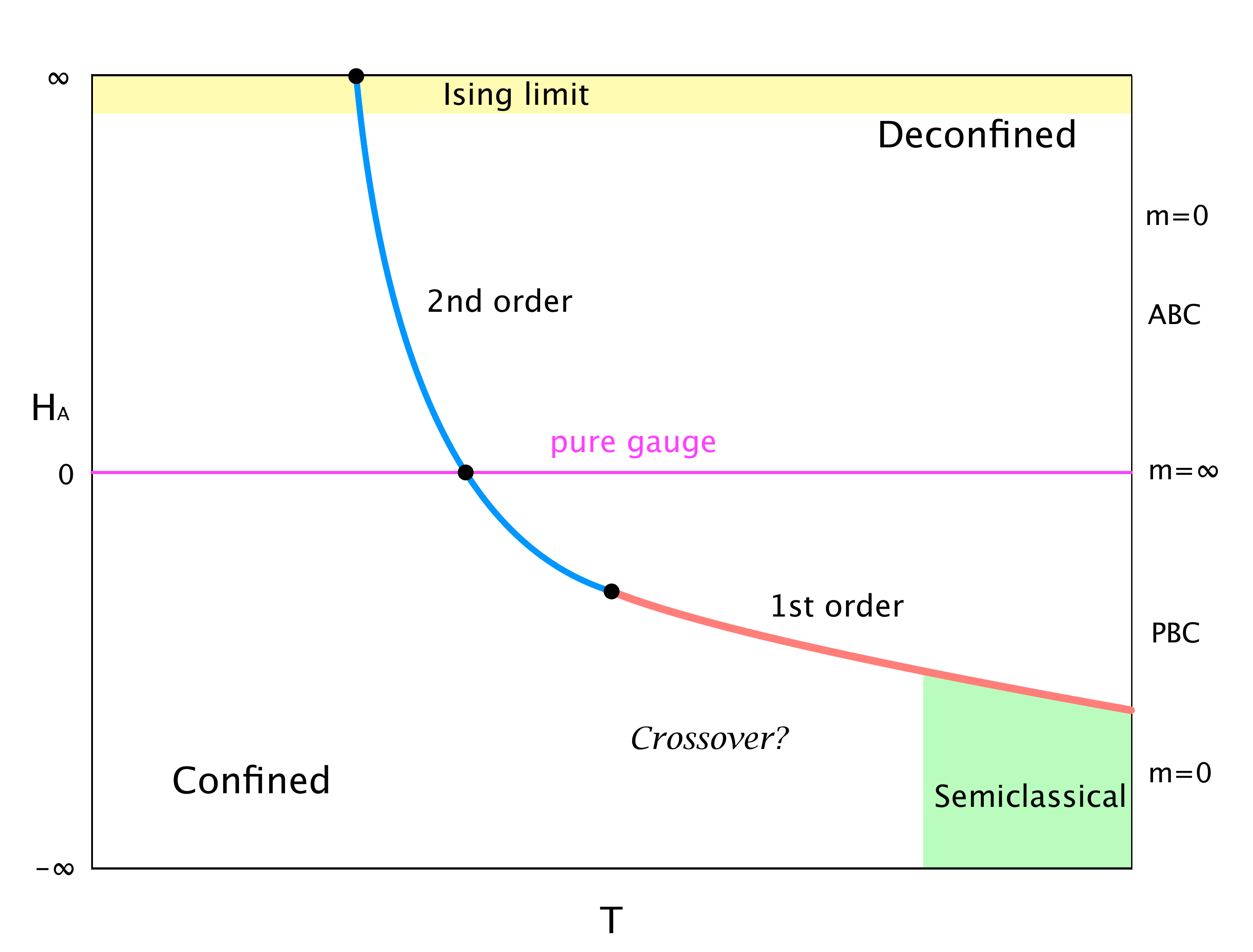}

\caption{\label{fig:su2-phase-diagram-1}$SU(2)$ phase diagram}
\end{figure}

The case of $SU(2)$ is particularly simple, because only two phases
are known: the confined phase and the deconfined. For $SU(2)$, the
double-trace deformation term added to the action can be taken to
be 
\begin{equation}
S\rightarrow S-\beta\int d^{3}x\, H_{A}Tr_{A}P\left(\vec{x},x_{4}\right)
\end{equation}
where we have now written $a_{1}$ to be $-\beta H_{A}$. If the coefficient
$H_{A}$ is sufficiently negative, the deformation will counteract
the effects of the one-loop effective potential, and $Z(N)$ symmetry
will hold for small $L$. The schematic form of the phase diagram
in the $T-H_{A}$ plane for an $SU(2)$ gauge theory with this deformation
is shown in Figure \ref{fig:su2-phase-diagram-1}. Positive values
of $H_{A}$ favor $Z(2)$ symmetry-breaking, and the critical temperature
will decrease as $H_{A}$ increases. In the limit $H_{A}\rightarrow\infty$,
the Polyakov loops will only take on values in $Z(2)$; this is therefore
appropriately described as an Ising limit. On the other hand, negative
values of $H_{A}$ favor $Tr_{F}P=0$. This leads to a rise in the
critical temperature. For the specific deformation considered here,
the critical line switches to first-order behavior at a tricritical
point. This behavior is familiar in $Z(2)$ models \cite{Nishimura:2011md}.
For sufficiently negative $H_{A}$, we reach the semiclassical region
where the running coupling $g(T)$ is small and semiclassical methods
may be applied reliably. Also shown in the figure is the rough equivalence
of a double-trace deformation with periodic adjoint fermions. Although
adjoint fermions couple in a more complicated way to the Polyakov
loop than the double-trace deformation, the two approaches are very
similar when the adjoint fermion mass $m$ is very large, and the
pure gauge theory is obtained in the limit $M\rightarrow\infty$.
Positive $H_{A}$ corresponds to conventional anti-periodic boundary
conditions while negative $H_{A}$ corresponds to periodic boundary
conditions. However, one cannot obtain the limits $H_{A}\rightarrow\pm\infty$
by taking $m=0$; this would require taking the number of adjoint
flavors $N_{f}$ to infinity, a limit incompatible with asymptotic
freedom. For fixed $N_{f}$, the $m=0$ corresponds to a finite value
of $H_{A}$, as shown on the right-hand axis.

\section{Monopoles, Instantons and Confinement on $R^{3}\times S^{1}$}

The non-perturbative dynamics of confining gauge theories on $R^{3}\times S^{1}$
are based on Polyakov's analysis of the Georgi-Glashow model in three
dimensions \cite{Polyakov:1976fu}. This is an $SU(2)$ gauge model
coupled to an adjoint Higgs scalar. The four-dimensional Georgi-Glashow
model is the standard example of a gauge theory with classical monopole
solutions when the Higgs expectation value is non-zero. These monopoles
make a non-perturbative contribution to the partition function $Z$.
In three dimensions, these monopoles are instantons. Polyakov showed
that a gas of such three-dimensional monopoles gives rise to non-perturbative
confinement in three dimensions, even though the theory appears to
be in a Higgs phase perturbatively. The models we are considering
thus differs by the addition of a fourth compact dimension and a change
to the action designed to maintain $Z(N)$ symmetry. 

\begin{figure}
\includegraphics[scale=0.6]{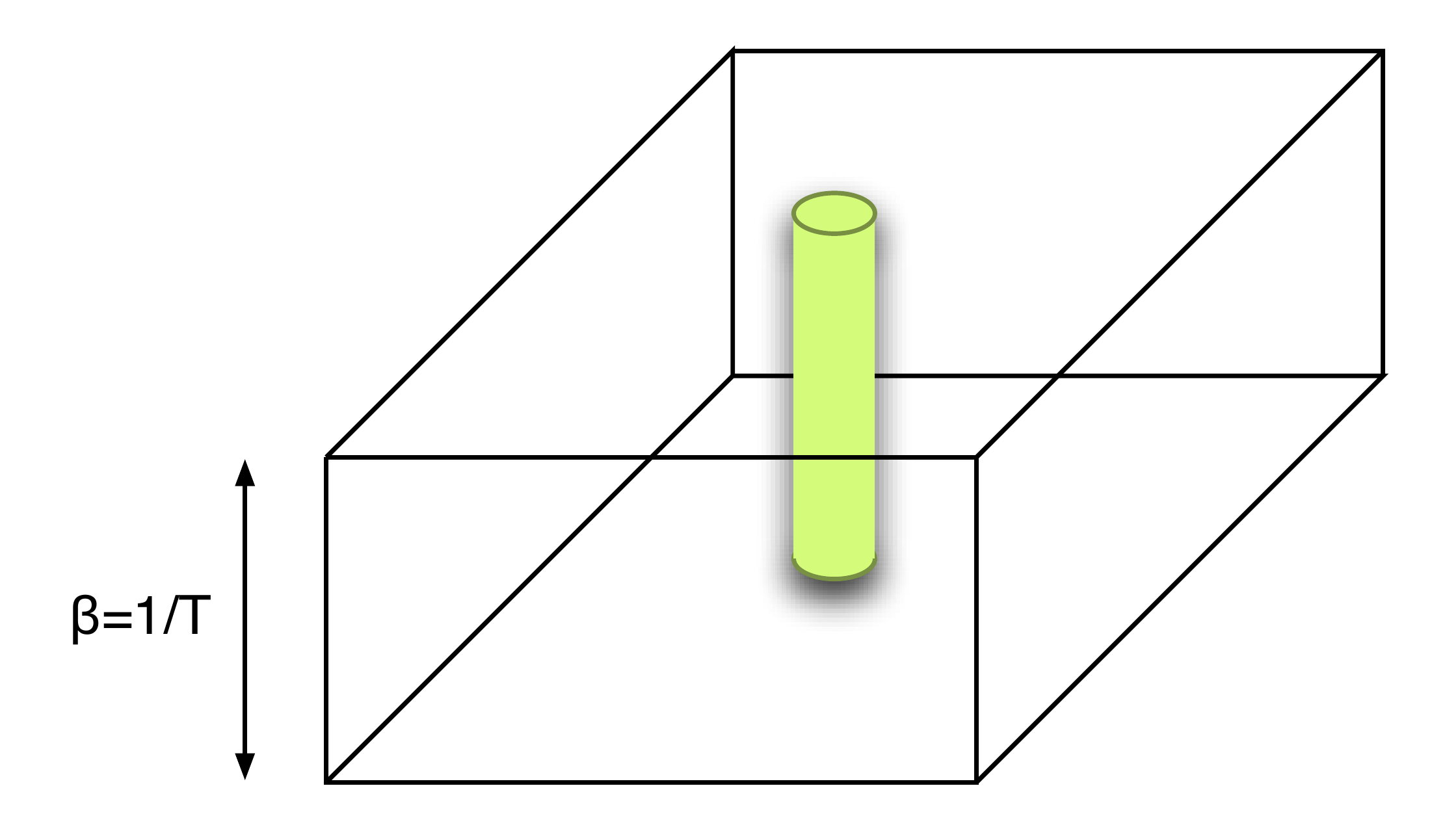}

\caption{\label{fig:Short-monopole-worldline}Short monopole worldline on $R^{3}\times S^{1}$.}
\end{figure}

Because $L$ is small in the $R^{3}\times S^{1}$ models we consider,
the three-dimensional effective theory describing the behavior of
Wilson loops in the non-compact directions will have many features
in common with the three-dimensional theory discussed by Polyakov.
In the four-dimensional theory, monopole solutions with short worldline
trajectories in the compact direction exist, and behave as three-dimensional
instantons in the effective theory; see Figure \ref{fig:Short-monopole-worldline}.
In models on $R^{3}\times S^{1}$, the role of the three-dimensional
scalar field is played by the fourth component of the gauge field
$A_{4}$. In a gauge where the Polyakov loop is diagonal and independent
of $x_{4}$, $P$ has a vacuum expected value induced by the perturbative
effective potential. However, there is another way to understand the
presence of monopoles in this phase, based on studies of instantons
in pure gauge theories at finite temperature and the properties of
the KvBLL caloron solution \cite{Lee:1998bb,Kraan:1998kp,Kraan:1998pm}.
If the Polyakov loop has a non-trivial expectation value, finite-temperature
instantons in $SU(N)$ may be decomposed into $N$ monopoles, and
the locations of the monopoles become parameters of the moduli space
of the instanton. In the case of $SU(2)$, an instanton may be decomposed
into a conventional BPS monopole and a so-called KK (Kaluza-Klein)
monopole. The presence of the KK monopole solution differentiates
the case of a gauge field at finite temperature from the case of an
adjoint scalar breaking $SU(N)$ to $U(1)^{N-1}$, in which case there
are $N-1$ fundamental monopoles. We will consider in detail the simplest
case of $N=2$.

The BPS monopole is found using the using the standard arguments \cite{Prasad:1975kr,Bogomolny:1975de}.
The Euclidean Lagrangian $\mathcal{L}$ can be taken to be 
\begin{equation}
\mathcal{L}=\frac{1}{4}\left(F_{\mu\nu}\right)^{2}++V_{eff}\left(P\right)
\end{equation}
where $V_{eff}$ includes both the one-loop gluonic effective potential
and the additional term that prevents $Z(N)$ symmetry breaking. This
can also be written as 
\begin{equation}
\mathcal{L}=\frac{1}{2}\left(D_{j}A_{4}\right)^{2}+\frac{1}{2}\left(B_{j}\right)^{2}+V_{eff}\left(P\right).
\end{equation}
We can associate with $\mathcal{L}$ an energy defined by 
\begin{equation}
E=\int d^{3}x\left[\frac{1}{2}\left(B_{j}\right)^{2}+\frac{1}{2}\left(D_{j}A_{4}\right)^{2}+V_{eff}\left(P\right)\right]
\end{equation}
as well as an action $S=LE$. We will concern ourselves for now with
the solutions in the BPS limit, in which the effective potential $V_{eff}$
is neglected, but the boundary condition on $P$ at infinity imposed
by the potential is retained. We can write the energy as 
\begin{eqnarray}
E & = & \int d^{3}x\left[\frac{1}{2}\left(B_{j}\pm D_{j}A_{4}\right)^{2}\mp B_{j}D_{j}A_{4}\right].
\end{eqnarray}
This expression is a sum of squares plus a term which can be converted
to a surface integral, giving rise to the BPS inequality 
\begin{equation}
E\geq\mp\int dS_{j}B_{j}A_{4}.
\end{equation}
The BPS inequality is saturated if the equality $B_{j}=\mp D_{j}A_{4}$
holds. For the case of a single monopole at the origin, we require
the fields at spatial infinity to behave as 
\begin{eqnarray}
\lim_{r\rightarrow\infty}A_{4}^{a} & = & w\frac{x^{a}}{r}\nonumber \\
\lim_{r\rightarrow\infty}A_{i}^{a} & = & \epsilon^{aij}\frac{x_{j}}{gr^{2}}.
\end{eqnarray}
Note that $w$ is related to the eigenvalues of $P$ at large distances
by $w=2\theta/gL$. Note that $A_{4}$ has the usual hedgehog form.
$A_{i}^{a}$ is chosen such that covariant terms vanish at infinity:
$\left(D_{i}A_{4}\right)^{a}=0$. With the 't Hooft-Polyakov ansatz,
the general expressions for the fields become 
\begin{eqnarray}
A_{4}^{a} & = & wh\left(r\right)\frac{x^{a}}{r}\nonumber \\
A_{i}^{a} & = & a\left(r\right)\epsilon^{aij}\frac{x_{j}}{gr^{2}}
\end{eqnarray}
where we define $w>0$ and require $h(\infty)=1$ or $-1,$ and $a(\infty)=1$
to obtain the correct asymptotic behavior. We must also have $h=a=0$
at $r=0$ to have well-defined functions at the origin. We identify
a magnetic flux 
\begin{equation}
\Phi=\pm\int dS_{j}B_{j}^{a}\frac{x^{a}}{r}=\mp\frac{4\pi}{g}
\end{equation}
where the $+$ sign corresponds to the case $h(\infty)=1$ and $-$
corresponds to $h(\infty)=-1$. The energy of the BPS monopole can
be written as 
\begin{eqnarray}
E_{BPS} & = & \mp\Phi w=\frac{4\pi w}{g}.
\end{eqnarray}

In addition to the BPS monopole, there is another, topologically distinct
monopole which occurs at finite temperature when $A_{4}$ is treated
as a Higgs field \cite{Davies:1999uw}. Starting from a static monopole
solution where $\left|A_{4}\right|=w$ at spatial infinity, we apply
a special gauge transformation 
\begin{equation}
U_{special}=\exp\left[-\frac{i\pi x_{4}}{L}\tau^{3}\right]
\end{equation}
 where $\tau^{i}$ is the Pauli matrix. $U_{special}$ transforms
$A_{\mu}$ in such a way that the value of $A_{4}$ at spatial infinity
is shifted: $w\rightarrow w-2\pi/gL$. If we instead start from a
static monopole solution such that $A_{4}=2\pi/gL-w$ at spatial infinity,
then the action of $U_{special}$ gives a monopole solution with $A_{4}=-w$
at spatial infinity. A final constant gauge transformation $U_{const}=\exp\left[i\pi\tau^{2}/2\right]$
yields a new monopole solution with $A_{4}=w$ at spatial infinity.
The distinction between the BPS solution, which is independent of
$x_{4}$, and the KK solution is made clear by consideration of the
topological charge. The action of $U_{special}$ followed by $U_{const}$
increases the topological charge by $1$ and changes the sign of the
monopole charge. Thus the KK solution is topologically distinct from
the BPS solution because it carries instanton number $1$. The $\overline{BPS}$
antimonopole has magnetic charge opposite to the BPS monopole, and
hence the same as that of the $KK$ monopole. The $\overline{KK}$
monopole has the same magnetic charge as the $BPS$ monopole, but
carries instanton number $-1$. This is all completely consistent
with the KvBLL decomposition of instantons in the pure gauge theory
with non-trivial Polyakov loop behavior, where $SU(2)$ instantons
can be decomposed into a BPS monopole and a KK monopole. Our picture
of the confined phase is one where instantons and anti-instantons
have ``melted'' into their constituent monopoles and anti-monopoles,
which effectively forms a three-dimensional gas of magnetic monopoles.
In the BPS limit, both the magnetic and scalar interactions are long-ranged;
this behavior appears prominently, for example, in the construction
of $N$-monopole solutions in the BPS limit. 

The BPS solution has action 
\begin{equation}
S_{BPS}=\frac{4\pi wL}{g}=\frac{8\pi\theta}{g^{2}}.
\end{equation}
 For the KK solution, we have instead 
\begin{equation}
S_{KK}=\frac{4\pi\left(2\pi-gLw\right)}{g^{2}}=\frac{4\pi\left(2\pi-2\theta\right)}{g^{2}}.
\end{equation}
The sum $S_{BPS}+S_{KK}$ is exactly $8\pi^{2}/g^{2}$, the action
of an instanton. For $\theta=\pi/2$, the $Z(2)$-symmetric value
for $SU(2),$ $S_{BPS}=S_{KK}$. This extends to $SU(N)$, where the
action of a monopole of any type is $8\pi^{2}/g^{2}N$ in the confining
phase. 

Although we used the BPS construction to exhibit the existence and
some properties of the monopole solutions of our system, we must move
away from from the BPS limit to ensure that magnetic interaction dominate
at large distances, \emph{i.e.}, that the three-dimensional scalar
interactions associated with $A_{4}$ and are not long-ranged. This
behavior is natural in the confined phase, where the characteristic scale
of the Debye (electric) screening mass associated with $A_{4}$ is
large, on the order of $g/L$. It is well known that the BPS bound
for the monopole mass holds as an equality only when the scalar potential
is taken to zero. Numerical studies \cite{Kirkman:1981ck} have shown
that the monopole action is given in general for $SU(2)$ as 
\begin{equation}
LE_{BPS}C\left(\epsilon\right)
\end{equation}
 where $C$ a function of the quartic term in the potential that varies
from $C=1$ in the BPS limit to a maximum value $C\left(\infty\right)=1.787$.
Thus corrections to the BPS result for the monopole mass and action
due to the potential terms are less than a factor of two. We will
henceforth use the exact results for the actions in the BPS limit,
neglecting corrections from $V_{eff}$ for the sake of simplicity
of notation. 

The $SU(2)$ construction of BPS and KK monopoles extends to $SU(N)$
in the standard way, via the embedding of $SU(2)$ subgroups in $SU(N)$.
There are $N-1$ BPS monopoles and 1 KK monopole inside an instanton.
In the confined phase, each of the $N$ monopoles has action $8\pi^{2}/g^{2}N$.
It has long been thought that instanton effects must be suppressed
in the large-$N$ limit, because instanton effects would vanish as
$\exp\left(-cN\right)$ in the limit $N\rightarrow\infty$ with $\lambda\equiv g^{2}N$
fixed \cite{Witten:1978bc}. In contrast, we see that the effects
of monopole constituents of instantons are not suppressed by the large-$N$
limit.

\begin{figure}
\includegraphics{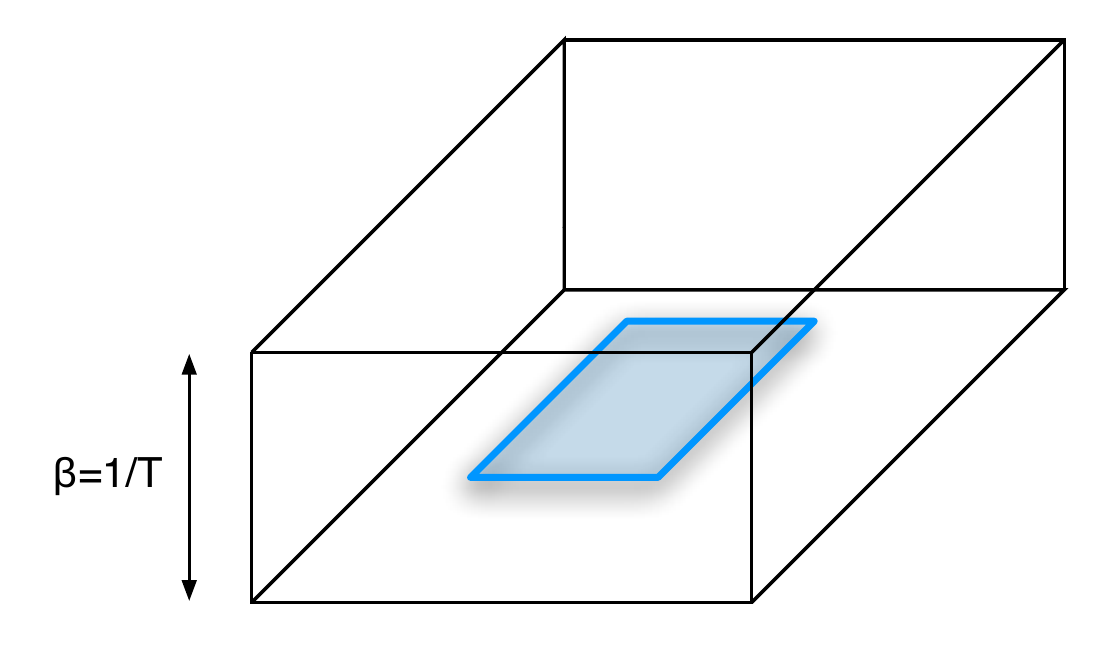}

\caption{\label{fig:Spatial-Wilson-loop}Spatial Wilson loop in $R^{3}\times S^{1}$
geometry.}
\end{figure}

In order to understand the effects of monopoles play in the confined
phase, we must analyze their interactions. We begin with a discussion
of quantum fluctuations around the monopole solutions. The contribution
to the partition function of a single BPS monopole at finite temperature
was considered by Zarembo \cite{Zarembo:1995am}. The measure factor
$d\mu^{a}$ associated with the collective coordinates (moduli) of
the monopole solution, including the Jacobians from the zero modes
is given by \cite{Davies:2000nw} 
\begin{equation}
\int d\mu^{a}=\mu^{4}\int\frac{d^{3}x}{\left(2\pi\right)^{3/2}}J_{x}\int_{0}^{2\pi}\frac{d\phi}{\left(2\pi\right)^{1/2}}J_{\phi}
\end{equation}
where $x$ is the position and $\phi$ the $U(1)$ phase of the monopole
and $\mu$ is a Pauli-Villars regulator. The label $a$ denotes the
type of monopole, $a=\left\{ BPS,KK,\overline{BPS},\overline{KK}\right\} $.
The Jacobians are 
\begin{equation}
J_{x}=S_{a}^{3/2},\,\, J_{\phi}=NLS_{a}^{1/2}.
\end{equation}
Each of the four zero modes contributes a factor of $\mu$. In the
BPS limit, each monopole carries an overall factor 
\begin{eqnarray}
Z{}_{a} & = & c\mu^{7/2}\left(NL\right)^{1/2}S_{a}^{2}\exp\left[-S_{a}+\mathcal{O}\left(1\right)\right]\int d^{3}x\nonumber \\
 & = & \xi_{a}\exp\left[-S_{a}\right]\int d^{3}x
\end{eqnarray}
in its contribution to $Z$ \cite{Zarembo:1995am}. The factor $\xi_{a}$
is $c\mu^{7/2}\left(NL\right)^{1/2}S_{a}^{2}$ where $c$ is a numerical
constant and the factor of $d^{3}x$ represents the integration over
the location of the monopole. From the construction of the KK monopole,
we see that we have $\xi_{KK}\left(\theta\right)=\xi_{BPS}\left(\pi-\theta\right)$.

The renormalization of the functional determinant arising from quantum
fluctuations around the monopole solution is particularly simple in
the confined phase, as first observed by Davies \emph{et al.} in the
corresponding supersymmetric model \cite{Davies:1999uw}. The dependence
on the Pauli-Villars regulator is removed, as usual, by coupling constant
renormalization. The relation at one loop of the bare coupling and
the regulator mass $\mu$ to a renormalization-group invariant scale
$\Lambda$ is 
\begin{equation}
\mbox{\ensuremath{\Lambda}}^{b_{0}}=\mu^{b_{0}}e^{-8\pi^{2}/g^{2}N}
\end{equation}
 where $b_{0}$ is the first coefficient of the $\beta$ function
divided by $N$: 
\begin{equation}
b_{0}=\frac{11}{3}-\frac{4}{3}\cdot\frac{n_{f}C(R_{f})}{N}-\frac{1}{6}\cdot\frac{n_{b}C(R_{b})}{N}
\end{equation}
 where $n_{f}$ is the number of flavors of Dirac fermions in a representation
$R_{f}$, $n_{b}$ is the number of flavors of real scalars in a representation
$R_{b}$, and $C(R)$ is obtained from $Tr_{R}\left(T^{a}T^{b}\right)=C(R)\delta^{ab}$.
For the case of a pure gauge theory with a deformation, there are
four collective coordinates and this gives a factor of $\mu^{4}$.
The functional integral over gauge degrees of freedom gives rise to
a factor $\det'\left[-D^{2}\right]^{-1}\propto\mu^{-1/3}$ and the
action contributes a factor $\exp\left(-8\pi^{2}/g^{2}N\right)$ in
the confined phase. Thus the contribution of a single monopole to
the partition function gives a factor 
\begin{equation}
\mu^{4-\frac{1}{3}}e^{-8\pi^{2}/g^{2}N}=\mu^{11/3}e^{-8\pi^{2}/g^{2}N}=\Lambda^{11/3}.
\end{equation}
Thus detailed calculation tells us that $\xi_{a}e^{-8\pi^{2}/g^{2}N}\propto L^{-3}\left(\Lambda L\right)^{11/3}$.
Note that the replacement of renormalization-dependent quantities
with renormalization-independent quantities depends crucially on the
coefficient of $1/g^{2}$ in the action. 

The interaction of the monopoles is essentially the one described
by Polyakov in his original treatment of the Georgi-Glashow model
in three dimensions \cite{Polyakov:1976fu}, generalized slightly
to include both the BPS and KK monopoles. Let us consider, say, a
BPS-type monopole and KK-type monopole located at $\vec{x}_{1}$ and
$\vec{x}_{2}$ in the non-compact directions, with static worldlines
in the compact direction. The interaction energy due to magnetic charge
of such a pair is 
\begin{equation}
E_{BPS-KK}=-\left(\frac{4\pi}{g}\right)^{2}\frac{1}{4\pi\left|\vec{x}_{1}-\vec{x}_{2}\right|}
\end{equation}
 and the associated action is approximately $S_{BPS}+S_{KK}+LE_{BPS-KK}$.
As discussed above, this will be larger than the value obtained from
the Bogomolny bound, but of the same order of magnitude. There is
an elegant way to capture the dynamics of the monopole plasma, using
an Abelian scalar field $\sigma$ dual to the magnetic field. Assuming
that the Abelian magnetic gauge field is three-dimensional for small
$L$, we may write 
\begin{equation}
L\int d^{3}x\,\frac{1}{2}B_{k}^{2}=\int d^{3}x\frac{g^{2}}{32\pi^{2}L}\left(\partial_{k}\sigma\right)^{2}
\end{equation}
 where the normalization of $\sigma$ is chosen to simplify the form
of the interaction terms. The three-dimensional effective action is
given by

\begin{equation}
L_{eff}=\frac{g^{2}}{32\pi^{2}L}\left(\partial_{j}\sigma\right)^{2}-\sum_{a}\xi_{a}e^{-S_{a}+iq_{a}\sigma}
\end{equation}
where the sum is over the set $\left\{ BPS,KK,\overline{BPS},\overline{KK}\right\} $.
Each species of monopole has its own magnetic charge sign $q_{a}=\pm$
as well as its own action $S_{a}$. The coefficients $\xi_{a}$ represent
the functional determinant associated with each kind of monopole,
but the combination $\xi_{a}\exp\left(-S_{a}\right)$ may be usefully
regarded as a monopole activity in terms of the statistical mechanics
of a gas of magnetic charges. The generating functional 
\begin{equation}
Z_{\sigma}=\int\left[d\sigma\right]\exp\left[-\int d^{3}x\, L_{eff}\right]
\end{equation}
is precisely equivalent to the generating function of the monopole
gas. This equivalence may be proved by expanding $Z_{\sigma}$ in
a power series in the $\xi_{a}$'s, and doing the functional integral
over $\sigma$ for each term of the expansion. 

The magnetic monopole plasma leads to confinement in three dimensions.
For our effective three-dimensional theory, any Wilson loop in a hyperplane
of fixed $x_{4}$, for example a Wilson loop in the $x_{1}-x_{2}$
plane, as shown in Figure \ref{fig:Spatial-Wilson-loop}, will show
an area law. The original procedure of Polyakov \cite{Polyakov:1976fu}
may be used to calculate the string tension, where the presence of
a large planar Wilson loop causes the dual field $\sigma$ to have
a discontinuity on the surface associated with the loop and a half-kink
profile on both sides. However, an alternative procedure is simpler
in which the discontinuity in the gauge field strength induced by
the Wilson loop is moved to infinity so that the string tension is
obtained from the kink solution connecting the two vacua of the dual
field $\sigma$ \cite{Unsal:2008ch}.

In the confined phase, the action and functional determinant factors
for all four types of monopoles are the same, so we denote them by
$S_{M}$ and $\xi_{M}$. The potential term in the confined phase
then reduces to 
\begin{equation}
-\sum_{a}\xi_{a}e^{-S_{a}+iq_{a}\sigma}\rightarrow4\xi_{M}e^{-S_{M}}\left[1-\cos\left(\sigma\right)\right]
\end{equation}
 which has minima at $\sigma=0$ and $\sigma=2\pi$; we have added
a constant for convenience such that the potential is non-negative
everywhere and zero at the minima. A one-dimensional soliton solution
$\sigma_{s}\left(z\right)$ connects the two vacua, and the string
tension$\sigma_{3d}$ for Wilson loops in the three non-compact directions
is given by 
\begin{equation}
\sigma_{3d}=\int_{-\infty}^{+\infty}dz\, L_{eff}\left(\sigma_{z}(z)\right)
\end{equation}
 which can be calculated via a Bogomolny inequality to be 
\begin{equation}
\sigma_{3d}=\frac{4g}{\pi}\sqrt{\frac{\xi_{M}}{L}e^{-S_{M}}}.
\end{equation}
This result depends on $L$ and is valid only in the region $\Lambda L\ll1$;
nevertheless, this is a concrete realization of confinement in a four-dimensional
field theory via non-Abelian monopoles.

It is possible to apply the same methods used for gauge theories on
$R^{3}\times S^{1}$ to other geometries, such as $S^{1}\times S^{3}$
or $R^{2}\times T^{2}$. Meyers and Hollowood have performed a detailed
study of $SU(N)$ gauge theories on $S^{1}\times S^{3}$ with periodic
adjoint fermions \cite{Hollowood:2009sy}. In this geometry, $R^{3}$
is replaced by $S^{3}$, so there are two length scales introduced
by the geometry, the radius of the three-sphere $R=R_{S^{3}}$ and
$L=R_{S^{1}}$. We require $\min\left[R_{S^{1}},R_{S^{3}}\right]\ll\Lambda$
so that we are in the weak-coupling region. The projection onto gauge-invariant
states, manifested as integration over the eigenvalues of the Polyakov
loop, ensures non-trivial behavior. Because the spatial volume is
finite, there is no actual phase transition for finite $N$, only
a crossover as $R/L$ is varied. However, the large-$N$ limit does
give a phase transition whose behavior is closely approximated even
for moderate values of $N$. These techniques can also be applied
to the study of gauge theories at finite temperature and density on
$S^{3}\times S^{1}$ \cite{Hands:2010zp,Hands:2010vw,Hollowood:2011ep}.
Another interesting geometry is $R^{2}\times T^{2}$, where $SU(N)$
gauge theories are in the universality class of $Z(N)\times Z(N)$
spin models because there are two compact directions \cite{Simic:2010sv,Anber:2011gn,Anber:2012ig}.

\section{Monopoles and Instantons on the Lattice}

\begin{figure}
\includegraphics{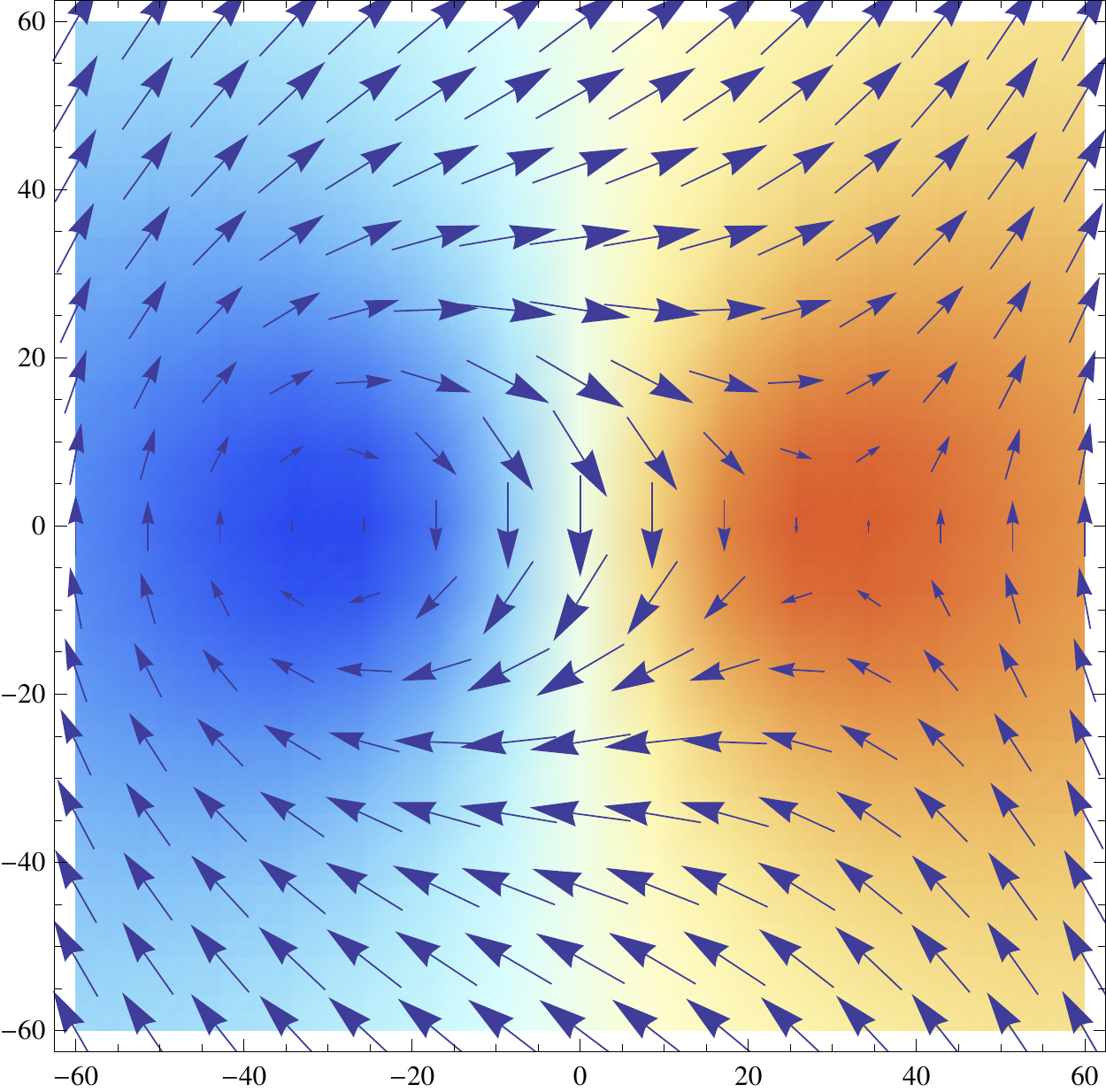}

\caption{\label{fig:Hedgehog-solution}An instanton solution of the $O(3)$
model showing its vortex-antivortex content. Arrows denote the field
components in the $\sigma_{1}-\sigma_{2}$ plane, and the colors denote
the value of $\sigma_{3}$.}
\end{figure}

The nonperturbative physics of confinement detailed in the previous
section is highly dependent on continuum methods that do not directly
extend to lattice models. It is illuminating to see how similar results
may be obtained from lattice models using rather different methods.
The key is the use of highly-developed techniques for Abelian lattice
duality. Before considering a four-dimensional $SU(2)$ lattice gauge
theory, we use the two-dimensional $O(3)$ model deformed to an $XY$model
to illustrate how lattice theories handle non-Abelian models deformed
to be Abelian.The continuum Euclidean action of the $O(3)$ model
is given by

\begin{equation}
S=\int d^{2}x\frac{1}{2g^{2}}\left(\nabla\vec{\sigma}\right)^{2}
\end{equation}
 where the field are constrained to 
\begin{equation}
\vec{\sigma}^{2}=\sigma_{1}^{2}+\sigma_{2}^{2}+\sigma_{3}^{2}=1
\end{equation}
 Like QCD, this is an asymptotically free theory that has instantons
\cite{Polyakov:1975rr}. As with finite temperature QCD, instantons
can be decomposed into constituents \cite{Gross:1977wu}. In the case
of the $O(3)$ model, these consitituents are XY-model vortices \cite{Ogilvie:1981yw}.
In Figure \ref{fig:Hedgehog-solution}, a classical instanton solution
is shown with the arrows denoting the components in the $\sigma_{1}-\sigma_{2}$
plane, and the colors denoting the value of $\sigma_{3}$. The embedding
of the vortex-antivortex solution within the instanton is obvious,
and $\sigma_{3}$ is near $\pm1$ precisely at the vortex cores. 

The $O(3)$ model can be deformed into an XY model by the addition
of a mass term for $\sigma_{3}$ \cite{Ogilvie:1981yw,Satija:1982ja,Affleck:1985jy}:
\begin{equation}
S\rightarrow S+\int d^{2}x\,\frac{1}{2}h\sigma_{3}^{2}
\end{equation}
in a manner similar to finite-temperature QCD. The mass term breaks
the classical conformal invariance of the model and makes it effectively
Abelian at large distances. It is physically obvious that as $h$
increases, the deformed $O(3)$ model will become more and more like
an XY model, and the constituent vortices inside instantons should
be identified with the Kosterlitz-Thouless vortices of the XY model.
To make these identifications precise, we consider a lattice form
of the deformed $O(3)$ model.

The lattice action is given by

\begin{equation}
S=-\sum_{x,\mu}K\sigma_{a}(x)\sigma_{a}(x+\mu)+\sum_{x}\frac{1}{2}h\sigma_{3}^{2}(x)
\end{equation}
 where $x$ is now a lattice site and $\mu$ one of two lattice direction;
the lattice parameter $K$ corresponds to $1/g^{2}$ in the continuum.
We parametrize $\vec{\sigma}$ as

\begin{equation}
\sigma=\left(\sqrt{1-\sigma_{3}^{2}}\cos\theta,\sqrt{1-\sigma_{3}^{2}}\sin\theta,\sigma_{3}\right).
\end{equation}
We can decompose the action as

\begin{equation}
S=-\sum_{x,\mu}K_{eff}\left(x,\mu\right)\cos\left[\theta\left(x\right)-\theta\left(x+\mu\right)\right]+S_{3}
\end{equation}
where
\begin{equation}
S_{3}=-\sum_{x,\mu}K\sigma_{3}(x)\sigma_{3}(x+\mu)+\sum_{x}\frac{1}{2}h\sigma_{3}^{2}(x)
\end{equation}
depends only on $\sigma_{3}$.

At this point, we can follow the well-known arguments of Jose {\it et al.}
\cite{Jose:1977gm} to obtain a form for the model that explicitly includes vortex
effects. This is done by a series of transformations on the Abelian
sector of the model. We write the partition function as
\begin{equation}
Z=\int_{S^{2}}\left[d\sigma\right]e^{-S}=\int_{-1}^{+1}\left[d\sigma_{3}\left(x\right)\right]e^{-S_{3}}\int_{S^{1}}\left[d\theta\right]\prod_{x,\mu}e^{K_{eff}\left(x,\mu\right)\cos\left(\nabla_{\mu}\theta\left(x\right)\right)}
\end{equation}
where $\nabla_{\mu}\theta\left(x\right)\equiv\theta(x+\mu)-\theta(x)$.
For each link, we expand the interaction in a character expansion,
which is a Fourier series: 
\begin{equation}
Z=\int_{-1}^{+1}\left[d\sigma_{3}\left(x\right)\right]e^{S_{3}}\int_{S^{1}}\left[d\theta\right]\prod_{x,\mu}\sum_{n_{\mu}\left(x\right)\in Z}I_{n_{\mu}\left(x\right)}(K_{eff}\left(x,\mu\right))e^{in_{\mu}\left(x\right)\nabla_{\mu}\theta\left(x\right)}
\end{equation}
where $I_{n}$ is a modified Bessel function. This step introduces
integer variables $n_{\mu}\left(x\right)$ on every link. We now make
use of the asymptotic form of $I_{n}$ for $K_{eff}\gg1$, using what
is called the Villain approximation, obtaining

\begin{equation}
Z=\int_{-1}^{+1}\left[d\sigma_{3}\left(x\right)\right]e^{S_{3}}\int_{S^{1}}\left[d\theta\right]\prod_{x,\mu}\sum_{n_{\mu}\left(x\right)\in Z}\frac{1}{\sqrt{2\pi K_{eff}\left(x,\mu\right)}}e^{K_{eff}\left(x,\mu\right)-n_{\mu}^{2}\left(x\right)/2K_{eff}\left(x,\mu\right)}e^{in_{\mu}\left(x\right)\nabla_{\mu}\theta\left(x\right)}
\end{equation}
Although this step appears here as an approximation, it is really
a small deformation of the action that does not change the critical
properties of the model. It is now easy to integrate over the $\theta$
variables, which leads to the constraint$ $$\nabla_{\mu}n_{\mu}\left(x\right)=0$.
This in turns allows us to write $n_{\mu}(x)=\epsilon_{\mu\nu}\nabla_{\nu}m(X)$
where $m\left(X\right)$ is an integer-valued field on the dual lattice
site $X$ which is displaced from $x$ by half a lattice spacing in
each direction. The partition function is now
\[
\]
\begin{equation}
Z=\int_{-1}^{+1}\left[d\sigma_{3}\left(x\right)\right]e^{-S'_{3}}\sum_{\{m\left(X\right)\}\in Z}e^{-\sum_{X,\nu}\left(\nabla_{\nu}m\left(X\right)\right)^{2}/2K_{eff}\left(x,\mu\right)}
\end{equation}
where
\begin{equation}
S'_{3}=S_{3}-\sum_{x,\mu}\left[K_{eff}\left(x,\mu\right)-\frac{1}{2}\log\left(2\pi K_{eff}\left(x,\mu\right)\right)\right]
\end{equation}
The final step is to introduce a new field $\phi(x)\in R$ using a
periodic $\delta$-function, effectively performing a Poisson resummation:
\begin{equation}
Z=\int_{-1}^{+1}\left[d\sigma_{3}\left(x\right)\right]e^{S'_{3}}\int_{R}\left[d\phi\left(X\right)\right]e^{-\sum_{X,\nu}\left(\nabla_{\nu}\phi\left(X\right)\right)^{2}/2K_{eff}\left(x,\mu\right)}\sum_{\{m\left(X\right)\}\in Z}e^{2\pi im\left(X\right)\phi\left(X\right)}.
\end{equation}

We see from this form of the partition function that vortices are
explicitly present in the functional integral, induced by the source $m(X)$
on the dual lattice. For each configuration $\{m\left(X\right)\}$,
the integral over $\phi$ and $\sigma_{3}$ must be carried out. This
can be done using standard perturbative methods. Each dual lattice
site $X$ where $m(X)\ne0$, will be the site of a vortex of charge
$m(X)$. In a dilute gas approximation, we can see that the size of
the vortex core will in general be set by the scale-setting parameter
$h$, which determines the region around $X$ where $\sigma_{3}$
is significantly different from zero. The contribution of the vortex
core to the total weight of a given configuration $\{m\left(X\right)\}$
can be captured in a vortex activity $y$, which represents the Boltmann
weight of the classical vortex solution times a functional determinant
factor, just as in the continuum. It is clear that in the limit where
$h$ is very large, $\sigma_{3}$ will be essentially zero everywhere,
and we recover the $XY$ model with $K_{eff}\simeq K$ and a vortex
core size on the order of the lattice spacing. Note that the $Z(2)$
symmetry under $\sigma_{3}\rightarrow-\sigma_{3}$ means that for
each vortex winding number $m$, there are two types of vortices depending
on the behavior of $\sigma_{3}$ in the core, as in the continuum.
For $h>0$, the large-distance behavior is that of an XY model, giving
a continuous path between the $O(3)$ model and the vortex Coulomb
gas phase of the XY model. If we keep only the $m=1$ contributions,
we have essentially a lattice sine-Gordon model
\begin{equation}
Z=\int_{R}\left[d\phi\left(X\right)\right]\exp\left[-\sum_{X,\mu}\frac{1}{2\bar{K}_{eff}}\left(\nabla_{\mu}\phi\left(X\right)\right)^{2}+\sum_{X}4y\cos\left(2\pi\phi\left(X\right)\right)\right]
\end{equation}
where $\bar{K}_{eff}$ is the value of $K_{eff}$ away from the vortex
cores. All of the physics associated with the shortshort-ranged $\sigma_{3}$
field is contained in $\bar{K}_{eff}$ and $y$.

The $(3+1)$-dimensional $SU(2)$ gauge theory at high temperatures
can be treated in much the same way as the two-dimensional $O(3)$
model \cite{Ogilvie:2012fe}. It is convenient to work in Polyakov
gauge, where $A_{4}$ is diagonal and time-independent so that the
Polyakov loop is given by $P=\exp\left(iA_{4}/T\right)=\exp\left(i\theta\tau_{3}\right)$.
Working at high temperature ensures that the running coupling constant
is small. A sufficiently strong deformation term will make the expected
value of the timelike link variable $U_{4}=\exp(iA_{4})$ significantly
different from one. This in term will give large masses to the off-diagonal
parts of the $U_{j}$ fields. The off-diagonal fields will be important
only inside monopole cores where $A_{4}$ is small. Outside monopole
cores, the model is effectively Abelian. 

A simplified approach is to take the deformation term to be very strong
and assume that all the fields are independent of $x_{4}$. We initially
take the the timelike links $U_{4}\left(\vec{x,}t\right)$ to be diagonal
and independent of $t$:
\begin{equation}
U_{0}\left(\vec{x}\right)=\cos\left(\theta_{0}\left(\vec{x}\right)\right)+i\sigma_{3}\sin\left(\theta_{0}\left(\vec{x}\right)\right)
\end{equation}
A strong deformation term forces $\left\langle Tr_{F}P\right\rangle =0$
with an expected value for $\left\langle \theta_{0}\right\rangle $,
given by $N_{t}\left\langle \theta_{0}\right\rangle =\pi/2$. As in
the $O(3)$ case, we can define the (dimensionally-reduced) spatial
gauge fields as 
\begin{equation}
U_{j}\left(\vec{x}\right)=\sqrt{1-\left(U_{j}^{1}\left(\vec{x}\right)\right)^{2}-\left(U_{j}^{2}\left(\vec{x}\right)\right)^{2}}\left[\cos\left(\theta_{j}\left(\vec{x}\right)\right)+i\sigma_{3}\sin\left(\theta_{j}\left(\vec{x}\right)\right)\right]+i\sigma_{1}\cdot U_{j}^{1}\left(\vec{x}\right)+i\sigma_{2}\cdot U_{j}^{2}\left(\vec{x}\right)
\end{equation}
The expectation value $\left\langle U_{0}\right\rangle $ makes the
$U_{j}^{1}$ and $U_{j}^{2}$ fields massive, and they do not contribute
to the large-distance behavior. This leaves us with an effective three-dimensional
$U(1)$ gauge theory. The dual of a a three-dimensional Abelian gauge
theory is an Abelian spin system, in this case again yielding a lattice
sine-Gordon model as in the continuum \cite{Banks:1977cc}.

The above simplified approach, based on the early application of dimensional
reduction, is in fact too simple. As in the $O(3)$ model, where there
were two types of vortices and two types of antivortices distinguished
by their behavior in the vortex core, there are four Eucldean monopole
solutions, not two \cite{Kraan:1998kp,Kraan:1998pm,Lee:1998bb,Davies:1999uw}.
The BPS-type monopole and anti-monopole solutions can be constructed
as conventional time-independent monopole solutions, and are thus
included in the simplified approach. On the other hand, the KK-type
solutions are constructed from the BPS solutions using an $x_{4}$-dependent,
non-periodic gauge transformation that changes the instanton charge
of a field configuration\cite{Davies:1999uw}. Thus, a proper treatment
of both types of monopoles is necessary. After accounting carefully
for both types of solutions, the dual form of the partition function
in the confined phase, reduced to three dimensions, is
\begin{equation}
Z=\int_{R}\left[d\sigma\left(X\right)\right]\exp\left[-\sum_{X,\mu}\frac{g^{2}}{8N_{t}}\left(\nabla_{\mu}\sigma\left(X\right)\right)^{2}+\sum_{X}4y\cos\left(2\pi\sigma\left(X\right)\right)\right]
\end{equation}
which has the same form as the corresponding continuum result, where
the effective action has the form
\begin{equation}
S_{eff}=\int d^{3}x\left[\frac{g^{2}(T)T}{32\pi^{2}}\left(\partial_{j}\sigma\right)^{2}-4y\cos(\sigma)\right].
\end{equation}
The two results are equivalent after indentifying $T^{-1}$ with $N_{t}$
and rescaling the $\sigma$ field. However, there are some significant
difference between the continuum approach and the lattice approach.
In the continuum, the renormalization group structure is more apparent.
In the lattice calculation, no Bogomolny bound is available for the
monopole action, and the correct approach to the continuum limit must
emerge in the limit where the lattice spacing is taken to zero. On
the other hand, the lattice approach manifestly includes all possible
combinations of monopoles as contributions to the partition function.
In contrast, in the continuum calculation combinations like a $BPS$
and $\overline{BPS}$ monpole pair, which are not topologically stable,
are not naturally included as an instanton effect and must be included
by hand. Moreover, such a continuum configuration is unstable to annihilation
of the $BPS$ - $\overline{BPS}$ pair, an effect that is naturally
avoided on the lattice.

\section{Conclusions}

The reduction of non-Abelian models to Abelian effective models is
a powerful technique that can be justified under certain conditions,
\emph{i.e.}, a small compactification radius and a suitable modification
of the action. Under those conditions, a very appealing picture of
confinement emerges that has a clear role for monopoles, instantons,
the maximal Abelian subgroup and center symmetry. Nevertheless, there
remain many unanswered questions. Chief among them is the connection
between the small-$L$, center-symmetric model where confinement can
be demonstrated and the corresponding large-$L$ confining behavior
that we wish to understand. There are also several major technical
issues in the treatment of instantons and monopoles. Center symmetry
is crucial in the recovery of the correct renormalization group behavior
for the confining instanton-monopole plasma. This raises serious questions
about the interpretation of monopole contributions when center symmetry
is broken. Center symmetry is slightly broken in the low-temperature
phase of QCD with dynamical quarks, where the Polyakov loop expected
value is small but nonzero. Another important issue is the treatment
of field configurations that have topological content but are not
topologically stable. This problem, present since the discovery of
instantons, is crucial here. Recent work on resurgence theory \cite{Argyres:2012vv,Argyres:2012ka,Basar:2013eka,Dunne:2014bca}
is a promising approach to understanding the correct treatment of
non-perturbative contributions in quantum field theory.

\bibliographystyle{unsrtnat}
\bibliography{review}

\end{document}